# Haptic Situational Awareness Using Continuous Vibrotactile Sensations

Muhammad Aakash Khaliq, Hammad Munawar and Qasim Ali

*Abstract*—In this research, we have developed a haptic situational awareness device that presents users with directional cues through continuous vibrotactile sensations. Using the device, we present user studies on the effectiveness of a torso-mounted haptic display in enhancing human situational awareness, when visual/auditory senses are degraded. A haptic display has been developed which generates continuous cues using tactile illusions and programmed to generate navigation commands. Participants are given navigation tasks where they may receive commands through any of the seven sensory modality combinations vibrotactile only, visual only, audio only, visual+vibrotactile, visual+audio, vibrotactile+audio and vibrotactile+audio+visual. Experimental results on human volunteers show that vibrotactile sensations enable enhanced situational awareness during sensory distraction as compared to visual-audio cues. This work is aimed at developing a viable haptic display for vehicle drivers, aircraft pilots and humans performing ground navigation, which can guide the wearer in situations when their visual/auditory senses may be degraded/overloaded.

## I. Introduction

Haptics is the use of sense of touch to convey information to the brain. It is further divided into *tactile haptics* and *kinaesthetic haptics*, where tactile haptics deals with the pressure, shear or vibration felt on the skin. When humans interact with their environment, tactile haptic sensations are naturally generated and play a very important part in enhancing their situational awareness of the environment. These sensations can also be created artificially with specialized actuators and are sensed by the human skin through *mechanoreceptors*. Mechanoreceptors are further divided into two categories, *meissner corpuscles* and *pacinian corpuscles*. Meissner corpuscles lie in the glabrous skin (non hairy skin like the hands) and are sensitive to vibration frequencies ranging from 5 to 50 Hz. Pacinian corpuscles lie in the non glabrous skin (hairy skin) and are sensitive to vibration frequencies ranging from 40 to 400 Hz [1].

When humans are interacting with the environment (such as walking in crowded or low light conditions) or operating vehicles (such as aircraft or cars), automated alerts and directional cues serve to guide their attention towards the task in hand and help them avoid potential dangers. These alerts are usually delivered using audio/visual channels in the form of displays, warning lights, audio commands, beeps, tones etc. However, in certain situations (e.g. during hazardous/emergency situations) when a lot of information is presented through multiple methods, visual and audio sensory channels may degrade. In this condition, an additional channel of information like sense of touch can be effective in providing critical information to brain. Moreover, this method of giving information through vibrotactile sensation is silent and inconspicuous to nearby users, giving it a certain advantage over other methods. This is the motivation for developing a vibrotactile display that assists with situational awareness. This is supported by the *multiple resource theory* [2], which states that humans possess multiple fixed-capacity resources (known as sensory channels or sensory modalities) for processing information obtained from multiple sources (such as audio, visual and touch). Thus, when one resource is not available, information can be provided to the brain using other vacant resources. In this context, the term *multi-modal* is to depict multiple display sources (visual, audio, vibrotactile) used together.

Systems that enable the sense of touch to present information are known as *Tactile Displays* [3]–[7]. Tactile displays consist of specialized actuators mounted inside specialized clothing, head gear, gloves, bands or belts. These actuators enable the information to be presented to the wearer in the form of sensations which may be discrete or continuous. Generation of discrete sensations requires multiple vibrotactile actuators. On the other hand continuous sensations can be generated using minimum number of actuators by specific control of intensity and frequency of vibration and temporal order. Continuous sensations provide a more natural feel (when they are generated within the peak sensitivity region of mechanoreceptors) [8], [9].

In addition to the above mentioned sensations, a phenomenon known as *Vibrotactile / Tactile Illusions* can be generated [1]. It is possible to trick the human body to perceive touch sensations that are different than what are actually being generated. Vibrotactile illusions occur on the skin when two actuators are activated in a coordinated pattern that their sensation is felt as combined instead of separate sensations. There are two types of tactile illusions, *Funneling Illusion* and *Sensory Saltation*. In the case of the *Funneling Illusion*, if a touch sensation is produced at two locations of a limb in a specified sequence, human will not feel any of these touch sensations, but feel a phantom sensation in between the two. It is also possible to control the exact location of this phantom sensation. Location of the phantom is controlled by varying the intensity and temporal order of vibrations [9]. Similarly, in the *Sensory Saltation / Cutaneous Rabbit* series of short vibration pulses are successively rendered at discrete locations on the

Muhammad Aakash Khaliq, Hammad Munawar and Qasim Ali are with the Department of Avionics Engineering, National University of Sciences and Technology, Islamabad, Pakistan
e-mail:a.khaliq,h.munawar,qasim.ali@cae.nust.edu.pk

skin, and the user feels as though a rabbit is hopping on the skin. In order to successfully generate haptic illusions, precise knowledge of human biology, electronics and programming is required [8].

Multiple actuator technologies have been used in creating vibrotactile displays such as *Linear Electromagnetic* actuators, *Rotary Electromagnetic* actuators and *Non Electromagnetic* actuators. C-2 actuator, Haptuator Mark-II, Linear Resonant Actuator (LRAs) are examples of linear electromagnetic actuators, Eccentric Rotating Mass (ERM) motors are examples of rotary electromagnetic actuators and Shape Memory Alloys (SMAs) are examples of non electromagnetic actuators [10], [11]. Details of the actuators have been discussed in hardware implementation section. These actuators are known by different names including tactile actuators, tactors and vibrotactile actuators.

Vibrotactile displays find applications in many areas. Alarms and alerts are commonly being used in mobile phones that present some vibration patterns to draw the attention of the user [3], [4]. It is common experience that in a noisy environment like a football stadium, visual and auditory channel of a mobile phone user may be overloaded causing her/him to miss an audio alert. However, the chances of missing a vibration alert are significantly less.

More advanced information such as navigation cues can also be presented through wearable tactile devices that guide the user [6]. Direction information can also be provided to people with visual and hearing disabilities through vibrotactile sensation on their skin [12]. Dogs can also be trained to take vibrotactile commands instead of vocal commands from humans [13]. Vibrotactile devices for navigation and obstacle avoidance of disabled persons sitting on power wheelchair have also been developed [14]. Similarly, vibrotactile cues find applications in challenging environments such as flying an aircraft, where warning messages must be robust and clear so that attention of the pilot could be drawn towards important information. Same is the case for landing of (helicopters) in degraded visual environment [7]. In addition to that, collision avoidance information can also be presented to astronauts when they have to perform space walks [5].

In this paper, our research contributes towards assessing the effectiveness of a torso mounted haptic display, for providing situational awareness information to users whose audio/visual senses may be degraded or distracted. The work also includes developing a low cost wearable vibrotactile display with commercially available vibrotactile actuators for displaying complex information to human wearers through continuous vibrotactile sensations. The sensations have been generated by the human sensory illusion called *Funneling Illusion* [8]. Proposed device consists of a belt worn on the torso, a control box, flexible polyurethane foam for vibration isolation and vibrotactile actuators, as shown in Figure 1.

This paper has been arranged into 6 sections. Section II comprises related work from literature. In Section III hardware and implementation of the proposed device is discussed in detail. Section IV is about the human experimentation of the device. Results are described in Section V while conclusions are drawn in Section VI.

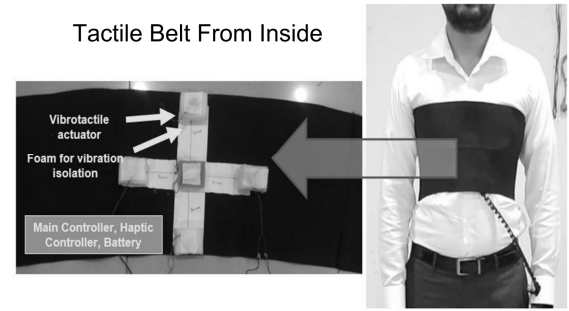

Figure 1: Proposed vibrotactile belt worn on the torso

## II. LITERATURE REVIEW

In this section, relevant work presented in literature has been reviewed. For comparison purposes, different bench marks have been selected including source of information to be encoded in terms of vibrotactile patterns, identification of coding parameters (vibration patterns and location of information presented on body) used in the existing devices, vibrotactile actuators used to generate vibration patterns, size of the array / numbers of actuators being used, application area and experimental results.

The most prominent system that uses tactile sensations for situational awareness is the Tactile Situational Awareness (TSAS) developed by the US Army Aero-Medical Research Laboratory (USAARL) in collaboration with NASA and more than a dozen other agencies and universities. Purpose of the system is to enhance situational awareness of pilots and has been tested on the T-34 jet aircraft and UH-60 helicopter. Recent efforts have been made to further develop and improve the TSAS. A version has also been developed for Special Forces [15], [16]. It is worn on the torso in the form of belts. Actuators used in the system are C-3 and eccentric mass rubber (EMR). System consists of 16 vibrotactile actuators integrated in a belt some of which are also integrated in the seat cushion. Vibrotactile actuators provide sense of touch in the form of vibrations and present specific flight parameters like direction, altitude and drift to the skin of pilot [17]. Hover information in degraded visual environment (DVE) can also be provided through a set of vibration intensities (no vibration at all, mild intensity, moderate intensity and strong intensity) [3]. Experimentation results on a UH-60 Black Hawk helicopter simulator showed that hovering near the selected location (in Brownout conditions) was three times better while wearing TSAS as compared to without using TSAS (relying only on visual and audio cues) [7]. A participant wearing the TSAS has been shown in the Figure 2. System presented in the study is aimed for specific demonstration of helicopter drift during hovering state i.e when helicopter drifts away from the intended hovering spot, then vibrotactile cues will guide the user (pilot) and will help the pilot remain over the intended hover area. Similarly, for degraded auditory conditions / noisy environment or people with hearing impairments, TSAS offers additional functionality of guiding the user to a certain direction by providing vibrotactile cues on the skin [7], [18].

Device presented by [19] has been developed at the Robotic

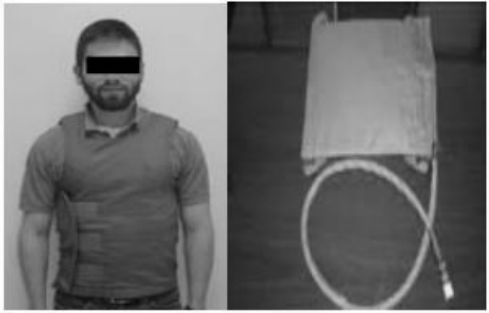

Figure 2: TSAS worn on the torso [7]

Sensor and Control Laboratory of South East University, China. It is a torso worn vest that provides aircraft roll and pitch information to the wearer in terms of various types of vibration patterns. Device is tested in a controlled laboratory environment on a simulator. Though vibrotactile vests and their different designs have been presented earlier in literature, a systematic approach about multiple coding parameters has been used in this device. For vibration delivery, coin type vibration motors have been used for their light weight, compact size, ease of amplitude adjustment and low price. Four columns of actuators (05 actuators in each column) have been integrated in a torso worn vest. Vibrotactile patterns have been generated by varying four basic parameters (location of the actuators, intensity of vibration, rhythm of vibration and mode of vibration). Pitch and roll information was presented through two columns of actuators in line with the frontal and sagittal plane of the torso respectively. A psycho-physical experiment was conducted in a static stimulating condition. Experimental set up has been shown in the Figure 3.

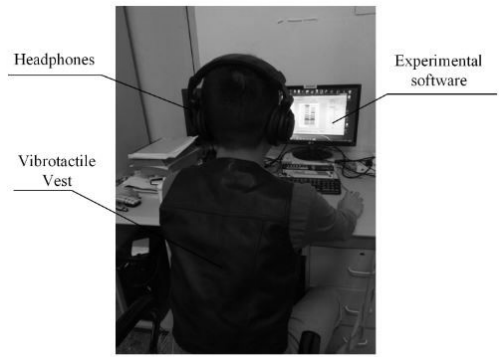

Figure 3: Experimental set-up of the device [19]

Objective of the experiment was to analyse the coding methods that were better understood by humans. Experiment consisted of independent variables (recognition accuracy, reaction time) and dependent variables (coding method, roll and pitch). Firstly a two parameter strategy followed by a three parameter strategy was used. Results of ANOVA showed that preferred coding method was a combination of location, rhythm and mode of vibration (LRM). The average success rate for all trials was around 95% for recognition accuracy and reaction time. So, it was concluded that three parameter strategy was better than two parameter strategy. Work presented in the paper is of worth in context of multiple parameters but it seems that it would take much training and cognitive effort to learn and be accustomed to the patterns being presented through vibrotactile vest. Moreover, flexibility in adjustment of actuator location could have been more effective in terms of stimulus perception.

Device presented by [20] has been developed at Complex Systems Control and Simulation Laboratory, Ecole Militaire Poly-technique, Algeria. Device has been used for flight envelope protection and enhancement of situational awareness of the pilot during sensory degradation / distraction. Device is equipped with a vibrotactile display that consists of an array of 16 ERM vibration motors. ERMs have been integrated inside a torso worn vest. In experimentation phase, experienced pilots tested the device on a flight simulator in a controlled laboratory environment. During experiments, information selected to be presented through vibration stimuli was flight path, climb angle, pitch limit, gamma limit and flight envelope limit. These vibration stimuli were delivered on front and back of the torso according to increasing and decreasing values of these parameters. Device has been shown in the Figure 4.

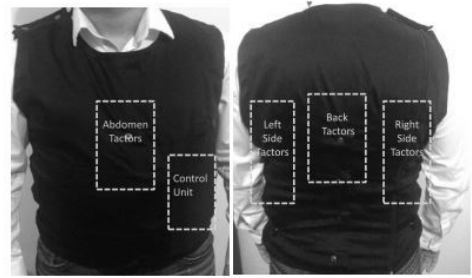

Figure 4: Vibrotactile device worn on the torso [20]

Vibration stimuli have been presented on left and right side of the torso in the form of very small (VS), small (S), medium (M), nominal big (NB) and very big (VB) frequency of vibration. During experiments three (03) experienced pilots executed five (05) flying scenarios. Objective of the experiment was to asses the significant differences between bimodal and visual only modes. Results achieved by one way ANOVA show that: there is a significant difference between bi-modal and visual only mode. Though presented work offers significant contribution but the change in frequency of vibration pattern may lead to confusion and may require considerable training time for the user to discreetly identify that which information is being presented.

The tactile cuing system presented by [21] has been developed at The Ohio State University. System presents in-flight icing information to pilots in the form of vibrotactile sensation. It is a forearm worn belt that consists of two vibrotactile actuators (vibro-tactors) located at a distance of 4" from each other. One actuator near the wrist is used to present in-flight icing formation information on the wing and second actuator placed near the elbow is used to present icing formation information on the tail of the aircraft. For vibration presentation, a frequency of 250 Hz is selected while 3 Hz of cycling vibration for light icing, 8 Hz cycling vibration for moderate icing and continuous vibration for severe icing. During experiments, a secondary task of monitoring the oil pressure was assigned through visual display and Air Traffic Control (ATC) instructions were being provided through auditory interface. Results show that there is no significant difference between modalities but with the addition of tactile display following factors such as error in missing the icing cue and incorrect icing identification have been reduced. Moreover, secondary task performance was also improved with the addition of vibrotactile display. Presented device has a low resolution of icing information moreover during performing the tasks in the cockpit, pilot's arm is continuously in use which may lower the chances of the device to be used in actual system. Preferably, the location of stimulus should be a fixed body part, not a moving one. This requirement makes the torso and thigh suitable candidates for being used as a location of vibrotactile display [15], [22].

Tactile Collision Avoidance System presented by [22] has been developed at LAHAV Division- Israel Aerospace Industries, Ltd. Aim of the system is to enhance the situational awareness of the pilots and to prevent the possibility of aircraft and hover craft crashes by providing tactile cues. Device consists of eight C-2 actuators which have been integrated on a thigh worn belt and the inter-actuator distance is 6 cm. Location of actuators has been shown in the Figure 5. Frequency of vibration was selected to be 250 Hz as it is the best perceivable frequency. Pattern of vibration was either continuous buzz of 250 Hz or pulsed buzzes. Three experiments were conducted. First one was to verify the suitability of vibrotactile belt on three different thigh positions from which belt positioned near knee gave better results. Second experiment was conducted to examine the added value of tactile directional information and the effect of message delivered by the vibrotactile belt. Results reported that vibrotactile-only display showed an improvement in reaction time of the participants while in the presence of visual cues, the effect of vibrotactile display was reduced. Also the stimulus produced by adjacent tactors was found to be confusing by users. In the third experiment, effect of cardinal sites was examined with four tactors as previously diagonal tactors were not playing a significant role and were creating confusion in perception of the stimuli. So, with four tactors it was observed that diagonal directions produced by two tactors did not help in enhancement of threshold perception. It only increased the accuracy of the direction. This device presented a novel location of information presentation on human thigh but thigh has a limited haptic area as compared to torso, as torso offers huge haptic space and a rich amount of mechanoreceptors lying in the torso skin [24], [25].

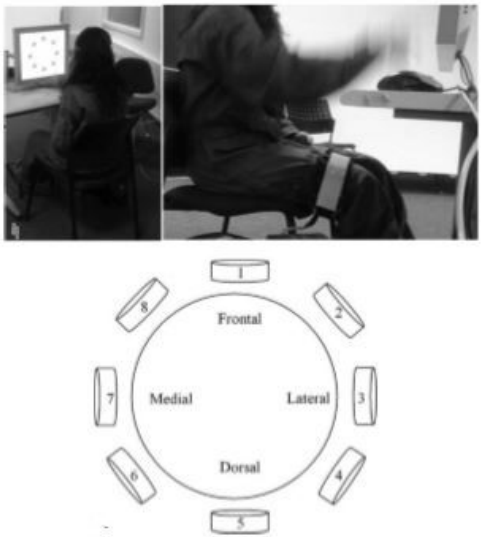

Figure 5: TCAS worn on thigh [22], [23]

Device presented by [13] has been developed at Ben-Gurion University of Negev. It is used for tactile communication between a human and a dog. Previously, only vocal commands were given to dogs to perform some tasks like turn around, lie down and walk backward. But with the help of this device a tactile vest that consists of four actuators is worn around the body of the dog and the dog is trained for multiple tactile cues. Four types of tactile cues have been presented in the form of pulsing and constant vibrations. After successful training, the dog can perform the tasks according to commands given through a remote control from the human. Commands like spin, down, to me and back pedal are given through tactile cues along with vocal cues. Experimental results show that dog missed 1 out of 18 commands with tactile cues and with vocal cues dog missed 3 commands out of 18 commands. Tactile cues gave promising result in communication between human and dog. This is quite a new idea of using tactile cues for communication with dogs but it needs more number of dog subjects for testing and also the testing on different breeds of the dogs. Moreover, generalization of the device is not presented as it is tested only on one subject dog. Tactile device worn on body of dog has been shown in the Figure 6.

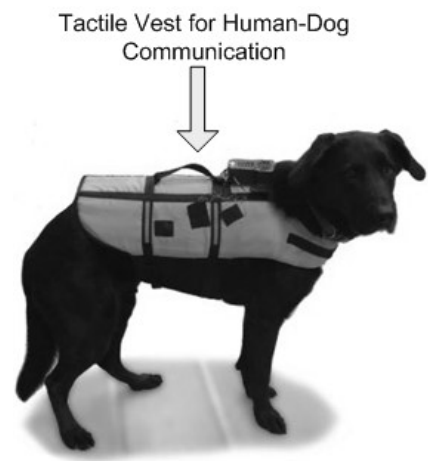

Figure 6: Tactile vest worn on the body of a German Shepherd Dog (test subject) [13]

Device presented by [26] has been developed under the EU FP-7 SME Program. Aim of the device is to present distance, obstacle avoidance and object ID information to the visually impaired people. Device is a torso worn vest consisting of 9 coin type vibration motors. Actuators have been integrated inside the belt in the form of a 3 by 3 array. Spatial, temporal and intensity variations were used for tactile patterns. Device was experimented for objectives like object distance, obstacle height, object direction and object ID. Task was to identify the object and categorize according to different categories of the objects. Each category of object has specific tactile cues associated to it. Results showed that system may help visually impaired people in navigation and for detection and avoidance of the obstacles along the path. Device has been shown the Figure 7. While it can be observed that this system may aid the visually impaired during navigation but its categorization of objects (according to height and distance) may cause too much information to be presented and may overwhelm the user.

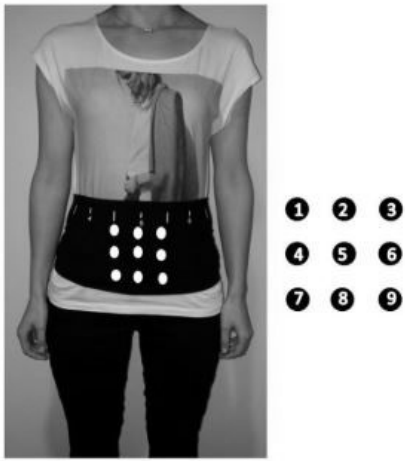

Figure 7: Participant wearing the vibrotactile vest [26]

Device presented by [27] has been developed at the University of Iceland. Objective of the device is to use spatio-temporal illusions to create upward and downward direction indication with minimum actuators. Purpose of using the illusion is to encode complex information by making use of less number of actuators. Device is in the form of a belt that consists of an array of 16 actuators. It is worn on the back of the torso. Actuators used in the device for vibration stimulus are Eccentric Rotating Mass (ERM). For information encoding parameter, two intensity of vibrations have been used: strong and weak. Device was experimented in the laboratory environment to verify the illusion generated by varying intensities (strong-weak and weak-strong) and inter-actuator distances (40 mm, 20 mm, zero). Two actuators were activated with the combination of weak and strong intensity of vibration to create the up and down responses. In the results, it was found out that intensity variation did create an error of localization and also the illusory motion effect was apparent with the inter-actuator distance of 20 mm and decreased with an increase in distance between the actuators to 40 mm. This new illusion may potentially decrease the size of tactile array and may lead to complex information encoding with minimum actuators but empirical exploration of the illusion is still required. Moreover, the effect of illusion in different conditions and body sites is yet to be explored. Also the numerical evaluation of frequency differences and optimum inter-stimulus interval should be specified. Device has been shown in Figure 8.

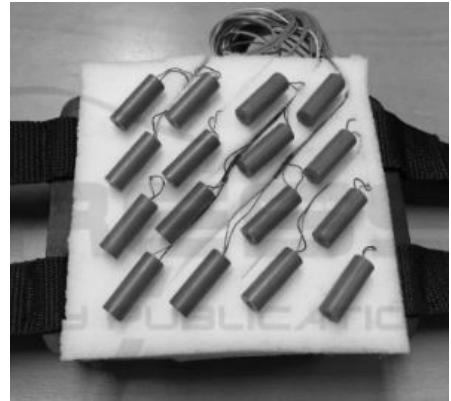

Figure 8: Device use new spatio-temporal illusion for up and down responses [27]

Device presented by [28] has been developed at the University of Ottawa. This work was supported by Natural Sciences and Engineering Research Council of Canada (NSERC). It was developed to asses the quality of continuous tactile perception for logarithmic and linear intensity of vibration. With intensity variations the Funneling illusion has been used to create continuous tactile perception. Device was in the form of armband straps that consists of four pan-cake DC motor type tactile actuators for creating vibration stimulus. Control factors for creating continuous tactile motion include Limb axis (transverse or longitudinal orientation), Duration of Stimulus (DOS), Gender (female, male), Limb site (forearm, upper arm) and Intensity variation (linear, logarithmic). Five Statistical tests (ANOVA) were performed to asses the control factors of continuous motion. Experiment was performed and data was collected to be used for statistical analysis. In the result, it was found out that two control conditions; Duration of Stimulus (DOS) and Intensity Variations have significant main effect on the perception quality of continuous tactile motion. Device has been shown in the Figure 9. This work has provided the parameters to be looked upon while creating continuous tactile motion but it lacks the level of generalization for different skin types and body sites. Moreover, we claim that continuous motion can be created with three actuators instead of four. So, the size of array can also be decreased with the proposed approach that we have used in the proposed device.

Device presented by [29] has been developed at the Graduate School of Engineering, Department of Robotics, Tohoku University, Sendai, Japan. It was is for navigation purposes. Device is worn on lower left leg near the ankle and 6 coin type vibration motors have been integrated inside an elastic strap. Developers used Phantom Tactile Sensation for presenting directional information to the user. Device was experimented by 15 volunteers for static and dynamic use and the reported results showed that there was a $15.36^0$ difference between the perceived and cued direction. It has been observed that location selected for device may be the reason behind the significant error range. Moreover, if ankle or lower leg part is used as a location for navigation device then many a times there may be difficulty in moving, as user would be taking the step and vibration may interrupt the normal way of walking and also could restrict the movement during running. Device worn on the lower leg has been shown in Figure 10.

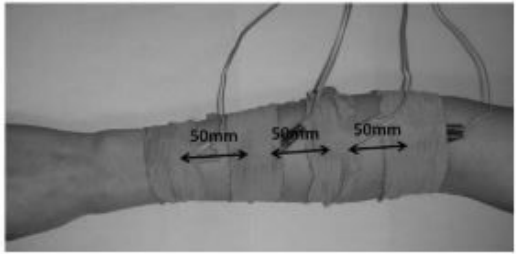

Figure 9: Armband straps worn on the forearm for creating illusory sensation [28]

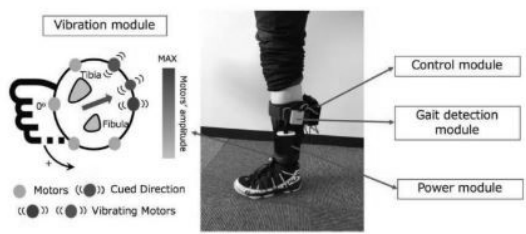

Figure 10: Participant wearing the haptic device on lower left leg [29]

Device presented by [14] has been developed under EU funded project "ADAPT" at the Pedestrian Accessibility Movement Environment, Laboratory (PAMELA), University College London (UCL). Device was used for navigation of power wheelchair for the people with severe disabilities. It was worn on the arm and consisted of two bands of four actuators each. Actuators were encapsulated in a housing and embedded on the elastic band and placed 90 degrees apart around the arm. Experimentation was done on fifteen healthy participants. When the wheelchair was close to an obstacle, actuators began to vibrate and guided the user to move in other direction. Results have shown that arm bands guided the user in multiple scenarios of collision avoidance and navigation. Though good results were reported but testing on people with actual disabilities would present more realistic scenario. A participant wearing arm band has been shown in Figure 11 .

Comparison of related devices that provide situational awareness information is shown in Table I. Though devices in literature provide continuous vibrotactile navigation information on legs [29] and hands [30], it is important to note that when user has to interact with the environment and perform different tasks, the legs and hands are not static. This would cause a hindrance in movement and reduce the effect of the vibrotactile display. Our proposed device is worn on torso, user can freely interact with the environment without any restriction, thus enhancing the effectiveness of the device.

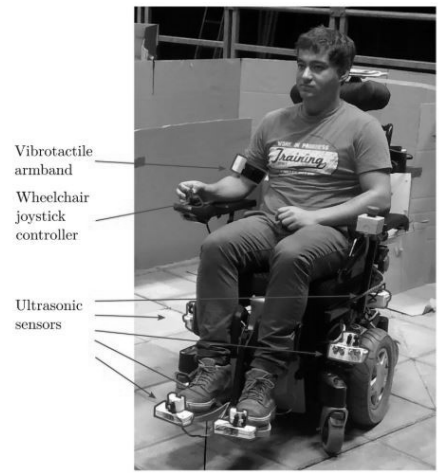

Figure 11: Participant is driving power wheelchair while wearing vibrotactile armband [14]

The device has then been experimentally verified on human volunteers.

## III. VIBROTACTILE ILLUSIONS

A phenomenon called *Sensory Saltation* was discovered back in 1970 at the Princeton Cutaneous Communication Laboratory while the word *saltation* means jumping in Latin. Initially three tactors were placed in line on the forearm of a person where the first one was placed close to the wrist. Then three very short pulses were given to the tactile actuators (tactors). Rather than just feeling the vibratory sensatiion on different locations of the forearm (locations where the tactors were placed), the subject felt as if a rabbit is hopping on the skin from the wrist towards the elbow. Hence, the phenomenon was named *Cutaneous Rabbit*. After the discovery of this phenomenon, extensive research has been done on cutaneous rabbit type sensations at Princeton University and it was observed by experiments that in order to create a rabbit like sensation the inter-tactor distance should not be greater then 10 cm [31]. Later on [32] studied the influence of vibrotactile illusions like *sensory saltation* and *cutaneous rabbit* and developed a tactile navigation system with a 3-by-3 tactor array worn on the back. For inter stimulus interval, a great effort has been made in a study by [33], where a judgement is made on the quality of continuous movement that *duration of stimulus (DOS)* is important in creating tactile illusions. Research continued and multiple types of illusions were discovered. They are now termed as *Vibrotactile Illusions / Tactile Illusions*, most common amongst which are the *Funneling Illusion* and *Phantom Sensation*.

Funneling Illusion happens on the skin when two actuators are activated in such a way that a virtual phantom appears on the skin midway between the actuators. Location of this phantom depends upon the relative vibration intensities of surrounding actuators and this location can be controlled by varying the vibration intensity of the actuators (which were used to create the phantom). When the intensity or temporal order, the virtual phantom feels like moving from one actuator towards the other. This is a well researched tactile illusion [8], [9], [28], [34]–[37]. Moreover, modification in the Funneling sensation can be done by two methods : *Temporal Inhibition and Amplitude Inhibition* [38], [39]. Temporal inhibition phenomenon happens when two discrete vibro-tactile actuators are activated with same intensities but at different time intervals. In result of this activity, funneled sensation is perceived to be moved toward the earlier activated actuator. This phenomenon has been shown in the Figure 12.

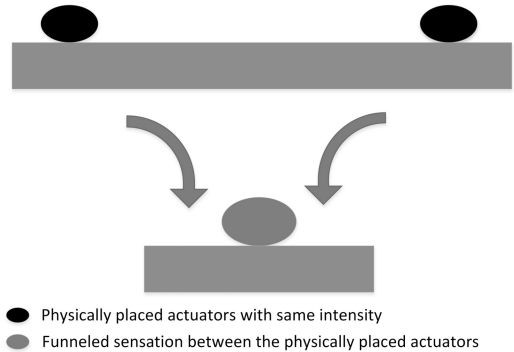

Figure 12: Funneling Illusion as a result of Temporal Inhibition

In case of amplitude inhibition, vibro-tactile actuators are activated with varying intensities but at the same time intervals. In result to this activity, funneled stimulus is perceived to be located near the actuator activated with higher intensity. This phenomenon has been shown in the figure 13.

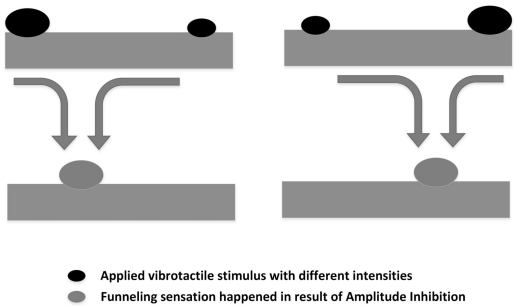

Figure 13: Funneling Illusion as a result of Amplitude Inhibition

The difference between Sensory Saltation and Funneling Illusion is such that in the former approach user feels discrete sensation like something is crawling on the skin. However, user feels discrete taps instead of continuous sensation [32]. On the other hand in case of Funneling illusion, by using amplitude and temporal inhibition, continuous sensations and variety of haptic / tactile effects are possible to create with good control over the parameters. With temporal inhibition virtual phantom appears on the skin between the physically placed vibrotactile actuators. Similarly, with the help of amplitude inhibition, generation of a continuously moving sensation is also possible [28].

In this paper, we have implemented the hybrid approach of using the amplitude and temporal inhibition. We activated the actuators at varying time intervals (a property of temporal inhibition) and used varying intensity of stimulus (a property of amplitude inhibition). By systematically applying the above mentioned technique, we are able to generate continuous moving sensation.

## IV. PROPOSED DEVICE (HARDWARE DESIGN AND IMPLEMENTATION))

In this section hardware design and implementation of the proposed device have been discussed. A vibrotactile display system consists of *Vibrotactile Actuators, Actuator Array, Vibration Pattern Generation and Information Encoding*. Therefore, each of sub systems for the proposed device are discussed in the following paragraphs.

### A. Vibrotactile Actuators

Vibrotactile sensation is artificially generated through vibrotactile actuators. Multiple types of actuators are available for this purpose. Some of them are custom designed for more reliability and robustness for specific application domains and others are more general in their use. Selection of a suitable vibrotactile actuator is important decision in designing a vibrotactile system. Two main hardware decisions are taken by the designer: firstly what type of actuator to use and secondly how to arrange them spatially so that vibrotactile sensation can be adequately felt by the user [10]. Examples of vibrotactile actuators available are Eccentric Rotating Mass (coin type), Linear Resonant Actuator (LRA), Eccentric Mass Rubber (EMR), Haptuator Mark 2 and C3 tactors. Shape of the above mentioned actuators has been shown in the Figure 14 and comparison based on features is shown in Table II.

In the proposed device, ERM is used for its affordability, availability, mechanical simplicity, electrical simplicity, moderate customizability and moderate expressiveness [20], [40]. ERM (model 310-118) by Precision Microdrives has a body diameter of 10 mm, operating voltage of 3-V, optimum vibration frequency of 240 Hz and a weight of 0.8 g that makes it suitable to be used in light wearable devices. ERM is also cost effective and readily available in the market. These features make the ERM the best candidate in comparison with other costly and complex actuators. In our design, we have used a frequency range of 200 to 250 Hz [33], [41] as it is best perceivable frequency range of mechanoreceptors in the torso region [24], [42].

*1) Biological Background of Human Vibrotactile Sensation:* According to biological background, vibration sensing capabilities of humans vary on different body sites and skin types [43]. Four sensory channel theory that identifies afferent neurons to perceive skin deformation, namely *Merkel Disks, Meissner Corpuscles, Pacinian Corpuscles and Ruffini Endings*, is considered as the base of vibrotactile perception. It asserts that: glabrous skin of humans carries all four types of mechanoreceptors [25], [44]. These four types of sensory channels respond to different set of mechanical stimuli that is summarized in the Table III. In non-glabrous skin Pacinian Corpuscles (PCs) and Rapidly Adapting (RA) mechanoreceptors are the reason behind vibration sensing while Meissner corpuscles are completely absent [1], [25], [40].

| Number | Authors | Developed by | Tactile Actuators | Tactile Array | Form factor | Location | Vibration Pattern | Encoded Information | Experimentation | Results |
|---|---|---|---|---|---|---|---|---|---|---|
| 1 | TSAS [15] | US ARMY / NAVY | C-3, EMR | 16 | Belt | Torso | Not made public | Hovering of helicopter | T-34 Jet aircraft, UH-60 Blackhawk | Three times better with TSAS |
| 2 | [19] | South-east University, China | DC motor | 20 | Vest | Torso | Intensity, rhythm, modes of vibrations | Roll & pitch | Flight simulator | 95% accuracy of rectangular flight patterns |
| 3 | [20] | Ecole Miitaire Polytechnique, Algeria | (ERM) | 16 | Vest | Torso | Not made public | Flight path climb, Turn rate | Flight simulator | Visual + vibrotactile better than visual only |
| 4 | [27] | University of Iceland | ERM | 16 | Belt | Lower back | Apparent shift of direction | Illusory upward or downward indication | Users indicated the direction of illusory stimulus | Activation of actuators aided in creating illusion |
| 5 | [28] | University of Ottawa | Pan-cake DC vibration motor | 4 | Armband | Arm | Linear and Logarithmic Intensity variation | Continuous motion on skin | Intensity and DOS varied to test the quality of sensation | Intensity variation and DOS has significant effect |
| 6 | [29] | Tohoku University, Japan | Coin Type Vibration Motor | 6 | Elastic Band | Lower Leg | Phantom Tactile Sensation | Collision alerts | Laboratory Environment | Perception of vibration decreased with fatigue |
| 7 | [14] | University College London (UCL) | Pico-Vibe | 4 - 4 | Armband | Arm | Phantom sensation | Navigation of power wheel chair | Laboratory Environment | Guided the user for obstacles. |

Table I: Comparison of most relevant situational awareness devices

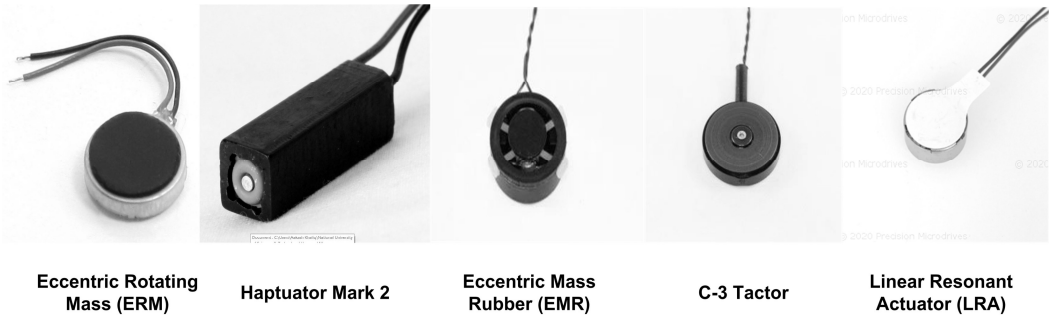

| | Eccentric Rotating Mass (ERM) | Haptuator Mark 2 | Eccentric Mass Rubber (EMR) | C-3 Tactor | Linear Resonant Actuator (LRA) |

Figure 14: Five commonly used vibrotactile actuators

| Features | ERM | Haptuator Mark 2 | EMR | C-3 | LRA |
|---|---|---|---|---|---|
| Company | Precision Microdrives | Tactile Labs | Engineering Acoustics | Engineering Acoustics | Precision Microdrives |
| Weight | 0.8 g | 9.5 g | 5 g | 8 g | 2 g |
| Dimensions | 10 × 2.1 | 9 × 9 × 32 mm | 25.4 × 10.2 mm | 20.3 × 6.35 mm | 10 × 3.7 mm |
| Optimum Freq. | 240 Hz | 120 Hz | 115 Hz | 240 Hz | 175 Hz |
| Operating Freq. | 160-280 Hz | 90-1000 Hz | 80-140 Hz | 180-320 Hz | N/A |
| Cost | 6 $ | 200 $ | 100 $ | 300 $ | 9 $ |
| Availability | High | Low | Low | Low | High |

Table II: Features of Vibrotactile Actuators

| Characteristics | Merkel Disks | Meissner Corpuscle | Pacinian Corpuscle | Ruffini Endings |
|---|---|---|---|---|
| Sensory Response | Slow Adapting | Rapid Adapting | Slow Adapting | Rapid Adapting |
| Skin Depth | Closer to skin surface | Closer to skin surface | Deep beneath surface | Deep Beneath Surface |
| Frequency Range (Hz) | < 5 | 3-100 | 10-500 | 15-400 |
| Perceived Stimulus | Pressure | Flutter | Skin Stretch | Vibration |
| Shape | Disk | Flattened stack of cells | Branched fibres in a capsule shape | Layered capsule shaped |
| Location | Between dermis and epidermis | Just below epidermis | Dermis | Deep in the skin |

Table III: Characteristics of Mechanoreceptors in Human skin [1], [10]

*2) Location of Stimulus:* From biological background of vibrotactile sensation it is evident that vibrotactile system designers should identify the suitable skin type and body location where vibrotactile stimulus is to be rendered. Skin type varies with body location so does the vibration sensing capability of the skin. Human hands and feet have glabrous skin which is most sensitive to even smaller frequency of vibration. While skin of the torso, legs, highs and arms have non-glabrous skin which is sensitive to higher vibration frequencies.

Location of vibrotactile stimulus rendering is chosen according to application. Since, our potential application is to provide navigation and directional information to pilots, visually impaired people and vehicle drivers so out of all body parts we have considered torso to be the suitable one. There are multiple advantages of using torso as location of vibration stimulus, because torso offers huge haptic space (which may be useful for increasing number of actuators in future) and it is mostly static (at least in the case of driving, piloting aircraft and for ground navigation of visually impaired people). Moreover, stimulus provided on the torso does not hinder movement of legs and hands which are already busy in doing other activities. So, torso is the best candidate for being used as location for vibrotactile stimulus. Furthermore, feasibility of selection of torso has also been discussed in literature [24], [42], [45]–[47].

*B. Array Design*

Proposed design is focused on presenting vibrotactile information on the torso as the torso offers a large haptic space and is mostly static during walking, driving, flying aircraft and operating machinery. Another key factor to consider while designing an array of vibrotactile actuators is the selection of a suitable vest, belt or elastic strap that could comfortably fit over the body part of the user. Garment selected for the belt or vest should be of elastic material to fit different torso sizes. Moreover, robustness of the belt during vigorous movements of the wearer and skin actuator contact is also fairly important. If skin-actuator contact is loose then vibration stimulation to the skin might be affected. Keeping in view the design requirements, a belt is selected to be used for wearing around the torso. Belt is made up of nylon webbing for flexibility and better wear-ability to any waist size. A square region of 170 by 170 mm is selected on the middle of the belt and an array of five actuators is drawn on the selected region with inter actuator distance of 70 mm [32], [46], [48]. To mount the actuators on the belt, foam of a specific grade and properties is selected to impede the vibrations from spreading elsewhere on the belt, to provide better skin-actuator contact and to present vibrations to the skin clearly. Array design has been shown in Figure 15.

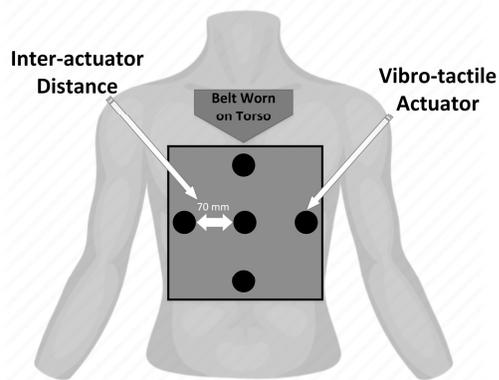

Figure 15: Proposed array of 05 actuators having an inter-actuator distance of 70 mm

Whereas, actual design created on belt has been shown in Figure 16.

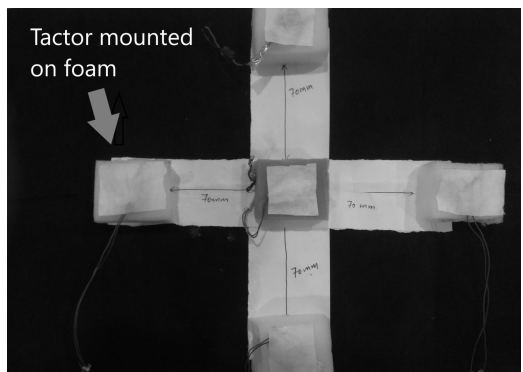

Figure 16: Array of proposed vibrotactile belt

## C. Vibration Patterns

For generation of vibration patterns with minimum delay a Control Box has been designed which consists of a micro controller, power supply and necessary circuitry. The circuitry ensures that sufficient starting current is available to vibrotactile actuators so that crisp and timely haptic effects can be generated. Each sub system and its operation/interface has been discussed in the paragraph below.

*1) **Control Box**:* The system is designed using an Arduino (NANO) open source hardware which is based on the Atmel Atmega 128P. It is the main controller that can be programmed through a PC. To address the problem of delays and generation of haptic waveforms, Immersion Corporation's Haptic Libraries were utilized [49]. Haptic driver (DRV-2605) made by Adafruit and originated by Texas Instruments has been used with the Arduino. Haptic driver 2605 forms the core of actuator module and controller [50], [51]. It supports both ERMs and LRAs having four libraries for ERMs and one library for LRAs. Furthermore, it provides flexible control of ERM and LRA over a shared $I^2C$ data bus and relieves the host processor from generating precise pulse-width modulated (PWM) signals. With the help of haptic driver, all the libraries can be accessed and used for different types of vibration waveforms.

Haptic driver is interfaced with micro-controller through data and clock pins. Micro-controller communicates with the haptic driver through $I^2C$ bus and sends data and clock signals. During design phase, reconfigurability of the device has also been considered in terms of change in vibration patterns and speed of vibration sensation.

*a) Data Communication Bus:* A very important requirement of the design of the proposed vibrotactile device is its reliability. Taking this requirement in to consideration, all the vibration motors have separate haptic controllers and communication between haptic controller and micro-controller occurs through $I^2C$.

*b) Power Source:* Considering the portability of the device for different applications, important physical and technical characteristics required for power source has been defined as small size, light weight, good backup time, safety in extreme weather and reliability. After analysis and consideration of physical and technical features an alkaline battery has been found suitable for the proposed vibrotactile device. Moreover, small size of alkaline battery provides a good charge density as compared to its size. Battery is rated at 9-V and provides almost 1.5 hour backup time for the continuous operation. Furthermore, it can be operated in a temperature range of $-18°C$ to $55°C$ [52].

*2) **Important Physical Parameters For Generating Tactile Illusions:**:* There are certain parameters that should be taken into account while creating vibrotactile sensations with vibrotactile actuators. For creating tactile illusions their importance further increases, because the parameters have to be precisely controlled within the range of vibration sensitivity. This range depends on the body location that has been selected for presenting tactile illusion cues. These parameters are shown below and each of them are discussed in the following paragraphs.

- Effect of Amplitude
- Frequency of Vibration
- Relation between Speed and Amplitude
- Relation Between Speed and Frequency
- Speed Variations

*a) Effect of Amplitude:* Practically, humans perceive the amplitude of vibration through our skin differently than that measured in controlled laboratory environment. So, vibration sensation felt through our skin may vary due to multiple factors like orientation of the motor, age of the user, location of the device on human body part (hand, arm, torso, leg), frequency of vibration and other environmental factors. Range of vibration amplitude is solely application specific like small device for children requires less amplitude. By changing the amplitude of vibration, different haptic effects can be created and this useful property of amplitude variations is being used to create different vibration patterns. Silent messages are conveyed to the user through different vibration patterns where each vibration pattern corresponds to a certain message or information.

*b) Frequency:* It is important to mention the relationship between amplitude and frequency of vibration. In the case of ERMs, amplitude and frequency of vibration are inseparable. Amplitude of the vibration is magnitude of the vibration waveform generated by ERM while frequency is the period of the vibration waveform. So, in the waveform as magnitude of the vibration increases, frequency of the waveform also increases. It is the difficulty that one quantity can not be changed without changing the other so most of the time, frequency is altered to alter the amplitude. Since we know that mechanoreceptors (Pacinian Corpuscle) in skin can perceive vibration frequencies ranging between 40 to 400 Hz while the peak sensitivity of vibration frequency occurs between 200 to 300 Hz [33], [41], [53], [54].

Moreover, another reason to use the peak sensitivity range of frequencies is due to the potential usage of the proposed device in automotive and aircraft environments. They have many sources of vibration in them like vibrations produced through running engine, air conditioner and road friction, that can also produce vibration at a certain range of frequencies. So, it is best to produce haptic effects at more distinguishable range of frequencies (200 to 300 Hz).

*c) Relation of Speed And Amplitude:* In any rotational system the relationship between speed and amplitude is exponential instead of linear. Centripetal force (F) produced by off centred mass in ERM can be described by the relation $F = mr\omega^2$. Here $m$ is the mass of eccentric load, $r$ is eccentricity and $\omega$ is rotational speed in radians/sec. A minor increase in motor speed will have a greater impact on the amplitude so we can make significant changes in the amplitude by just increasing motor speed.

*d) Relation Between Speed And Frequency:* Speed and frequency of ERM are one and the same thing. Motor speed is revolutions per minute and vibration frequency is represented in Hz. To convert from Hz to RPM we simply multiply by 60 and to convert from RPM to Hz we divide by 60.

*e) Speed Variations:* To change the speed of vibration motor we need to change the driving voltage, as driving voltage has direct effect on motor speed. Increase in driving voltage will increase the torque produced by the motor as the load is fixed motor speed will increase. For reliable motor operation and to get the good lifespan from motor it is often recommended by the manufactures to drive between two parameters. One parameter is maximum start voltage and other is maximum operating voltage. So, we need to stay within the constraints of voltage, because exposure to higher voltage for a long time may draw current beyond max rated current leading to motor failure.

*f) Amplitude And Frequency Prediction Difficulties:* It is difficult to predict the real vibration amplitude based on change in voltage because there is not a linear relationship between voltage and amplitude and it varies with different motors. So, we used the vibration motor performance graph to predict the vibration amplitude and frequency with respect to change in drive voltage that has been specified in the data sheet [55].

*3) **Creating Continuous Movement Sensation**:* We want to exploit the human sensory illusion called 'Funneling Effect' by which we will create continuous motion. When the vibration sensation is rendered on the human skin, he / she may feel as if something is moving on their skin [28]. For this purpose, we control the intensity of the adjacent actuators. We have two actuators separated by a distance of 70 mm which is the proposed inter-actuator distance [48], [56], [57].In this paper, we have implemented the hybrid approach of using the amplitude and temporal inhibition. We activated the actuators at different time intervals (a property of temporal inhibition) and used varying intensity of stimulus (a property of amplitude inhibition). By systematically applying the above mentioned technique, we are able to generate continuous moving sensation. Inverse proportional intensity technique of amplitude inhibition has been applied in such a way that when intensity of vibration of first actuator increases from low to high value then the intensity of second actuator will decrease from high to low intensity level. Therefore, the perceived continuous motion will move from the first stimulus position to the second stimulus position. At the same time temporal order of activation of actuators is also controlled in a specified way. This discrete intensity stimulus rendering on different body locations has been felt as continuously moving from one point to the other. By psychophysical influence of the Funneling illusion we successfully implemented the continuous tactile motion on the skin and have used this continuous tactile motion for presenting directional information.

Intensity variation technique is based on existing research, that by varying the intensity linearly or in a logarithmic way the stimulus can be felt between the actual stimulus locations. The work of [58], [59] describes the linear variation in stimulus amplitudes which causes the sensation to end near the mid point of both the stimuli.

*D. Encoding of Information*

Based on Funneling illusion (hybrid approach) we encoded the direction information in the device. So, device is capable of giving continuous directional vibration cues. Continuous sensation happening downward indicates *"Forward Movement"* as shown in Figure 17. Vibrotactile sensation occurring from bottom to top (upward) is aimed to command the user for *"Backward Movement"* as shown in Figure 18. In addition to that, continuous sensation rendered toward right side of the user indicates *"Right Turn"* as shown in Figure 19. Similarly, vibrotactile sensation rendered toward left side of the user signals the user to take *"Left Turn"* as shown in Figure 20. Moreover, vibro-tactile cue for *"Stop"* command is rendered in such a way that stimuli from four actuators approach the middle actuator as shown in Figure 21. Each directional vibration cue has been rendered on human skin with minimum of three actuators. Only three actuators have been used for direction information. For stop command, four actuators were activated.

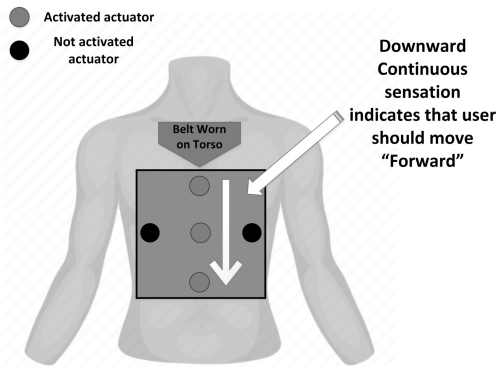

Figure 17: Continuous sensation is rendered from top actuator toward bottom actuator to indicate the cue of "Forward Movement"

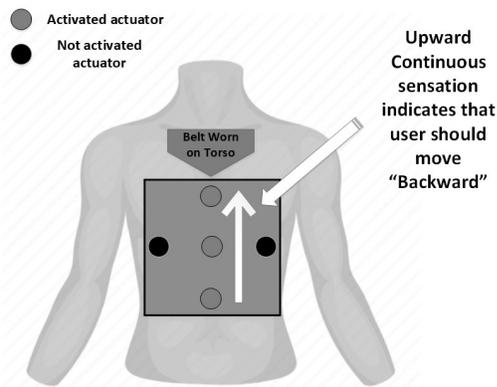

Figure 18: Continuous sensation occurring from bottom actuator toward top actuator signals "Backward Movement"

## V. EXPERIMENTATION

To evaluate the designed vibrotactile display, two important experiments have been performed. First experiment is designed to assess the perception of continuous vibrotactile sensations by users, while the second is designed to evaluate the effectiveness of the haptic display for ground navigation by users when their visual/auditory senses are degraded.

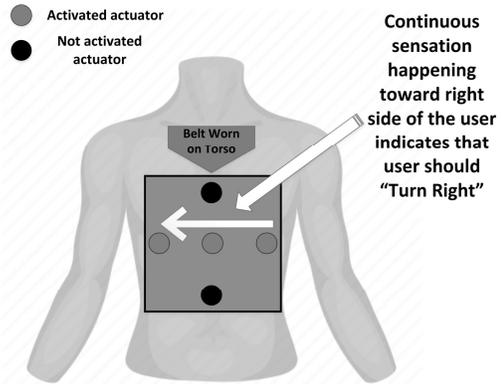

Figure 19: Continuous sensation toward the right side of the user commands the user to "Turn Right"

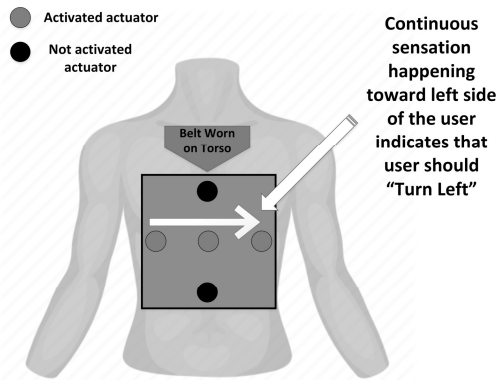

Figure 20: Continuous sensation stimuli happening toward left side of the user signals "Turn Left"

### A. Experiment Setup

15 healthy volunteer adults (10 males, 5 females) between the ages of 21 to 31 took part in the experiments. Informed written consent was obtained from the participants in line with the IRB's policy for Human Subjects in Research. Participants were ostracized from the experiment if they reported any physical injuries or cognitive impairments. All the subjects

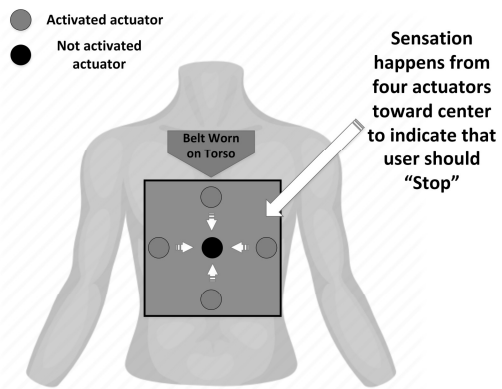

Figure 21: Stimuli approach from four actuators to the centred actuator indicate the "Stop" signal

were naive and did not have any previous knowledge of the vibrotactile device and its working.

For the second experiment, a navigation maze was created in a laboratory environment. Walls of the maze were made from mattresses, which not only made is safe for users but also enable easy configuration. Maze had several turns that user has to cross to reach the final destination. For each experiment the configuration of the maze was changed. A participant wearing vibrotactile belt in the navigation maze has been shown in the Figure 23.

Three *sensory channels* including vibrotactile, audio and visual have been considered. Individual sensory channels or their combinations have been considered *sensory modalities*. Modalities have then been grouped into *modes* which show how many sensory channels are involved. Modalities which consist of a single sensory channel form the *single-modality mode*. If two sensory channels are involved to present some information they are termed as *bimodal* and for more than two sensory channels *multi-modal* (it is the situation where most of the sensory degradation / distraction occurs). In this experiment each participant was presented with all of the sensory modalities as shown below.

- Vibrotactile only
- Visual only
- Audio only
- Visual + vibrotactile
- Visual + audio
- Vibrotactile + audio
- Vibrotactile + audio + visual

Vibrotactile cues were delivered through the proposed haptic device. Audio commands/distractions were given through bluetooth headphones, visual cues were given by making gestures and visual distractions were created by asking participants to play games on handheld devices and while navigating. Reason behind creating distractions was to assess the performance of vibrotactile display when visual and auditory sensory channels are degraded (vehicle drivers or aircraft pilots face this situation when dealing with emergencies). Another practical motivation behind creating the distraction was to impersonate a situation where a person is visually impaired, or a person who is using the phone (heads down) while walking across the road. Due to sensory degradation / distraction vehicle drivers, visually impaired persons and people who use phone during walk are likely to miss some important warning or information. For sensory modalities that did not require visual cues, participants were asked to wear black glasses or blindfolds depending on their preference.

### B. Data Collection and Analysis

During the experiment, data was collected and Statistical Analysis based on Three-Way ANOVA (analysis of variance) was performed to analyze the effect of *independent variables/categorical groups* on the *dependent variable*. Independent variables consisted of have designed four categorical groups including *gender* (two groups: male or female), sensory modality (seven: visual only, vibrotactile only, audio only, visual + audio, visual + vibrotactile, vibrotactile + audio and visual + audio + vibrotactile) and *reaction time* (in seconds: 1, 2, 3, 4). ANOVA helped in determining whether the different groups' mean scores differ from another or not [58], [59]

Certain assumptions were taken care of while performing ANOVA. First assumption states that dependent variable must be measured at a continuous level. In ANOVA the second assumption is that three independent variables should each consist of categorical groups and there should be at least two categorical groups [60], [61]. Third assumption states that all the data should be collected randomly and independently. Fourth assumption is about the normal distribution of the data. Fifth assumption asserts the importance of no significant outliers in the data. The problem with outliers is that they might have negative a impact on three way ANOVA [58]–[61]. All these assumptions were duly considered during our analysis.

### C. Assessment of Vibrotactile Sensations

Users wore the proposed device worn on the torso as in the Figure 22. Core objective was to asses the quality

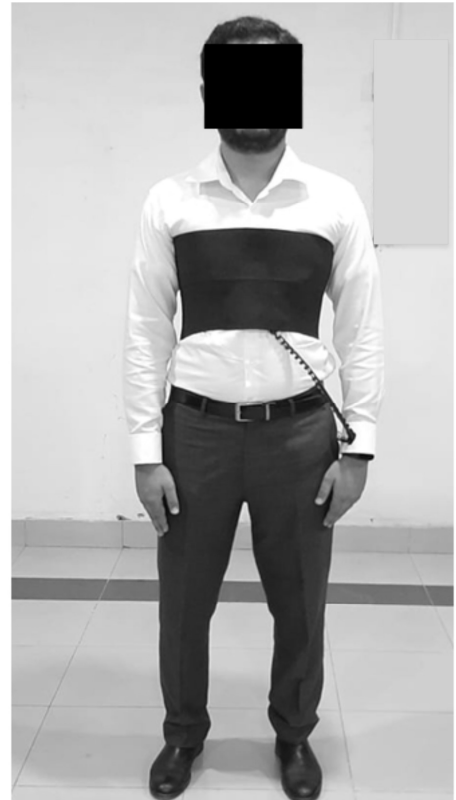

Figure 22: Proposed vibrotactile belt worn by a volunteer

and perception of continuous sensation on skin based on the feedback provided by the participants. In proposed device, the speed of continuous sensations has been kept programmable and termed as the *tactile motion speed*. It is directly related to the duration of stimulus of the continuous tactile motion in each direction. Tactile Motion Speed has two levels based on the duration of stimulus (fast - 450 ms and slow - 900 ms). We ran both levels (fast, slow) of motion speed parameter for each of the participants and took their feedback. Out of

15 participants (10 males, 5 females) only two males and 1 female wanted the sensation to be 'fast' while rest of the participants were comfortable with 'slow' level. During study, we wanted the participants to be comfortable with the vibrotactile sensory modality and get used to it. After the vibrotactile cues were validated successfully, we proceeded towards the second experiment.

*D. Effectiveness of Haptic Device During Ground Navigation*

In this experiment, users were tasked to complete the maze while receiving commands through the designated sensory modality. Configuration of the maze was changed before each experiment run.

First mode to be simulated in the experiment is the single modality. So, vibrotactile-only cues were presented to the participants for navigation through the maze, while participants were blind folded and distracting sounds were played through the headphones. Similarly, visual-only cues and audio-only cues were given, with the vibrotactile display switched off. Single sensory modality has been shown in Table IV. In the bimodal scenario, two sensory modalities were presented for ground navigation and third sensory modality was removed. Combination of bimodal display has been shown in Table IV. In the multi-modal, all the three sensory modalities (visual, audio, vibrotactile) were generated for the users.

Participants had to finish navigating the maze from start to end. They were not aware of the shape of the maze beforehand. Navigation maze had six check points / tasks, that user has to cross to reach the end point. Each of this check point / task carried a weight of "1" for successful check point and "0" for unsuccessful check point / task. Completion of all the check points / tasks rewarded the participant with 6 marks. All of the three modes of sensory modalities were used by the participants to perform the navigation maze experiment. In addition to that, in multi-modal combination, the visual/auditory senses of users were intentionally distracted along with the primary task of crossing navigation maze have been assigned to them. Furthermore, data of reaction time(time taken by the participant to perceive the direction) and task completion details of all of the participants for each of sensory modality were noted for further use in statistical analysis. A participant attending to visual display of a cell phone screen, while receiving vibrotactile navigation cues is shown in the Figure 24.

## VI. Results

ANOVA analysis showed some significant effects one the dependent variable (task completion) (Table V). Moreover, independent variables sensory modality and reaction time were found to have statistical significance while the independent variable gender is statistically insignificant as shown in (Table V).

*A. Effect of Sensory Modality on Task Completion*

Null hypothesis and Alternate hypothesis have been formulated to assess the *effect of Sensory Modality on Task Completion*. Null hypothesis states that, there is no significant effect of sensory modality on task completion if the condition ($p > 0.05$) is true for 95% Confidence Interval (CI) while on the other hand Alternate hypothesis states that there is significant effect of sensory modality on task completion if $p < 0.05$ for 95% Confidence Interval (CI). Trials depicted a significant main effect for the type of sensory modality (Table V), most prominently, when sensory degradation / distraction scenario was simulated. Positive effects have been observed when there was distraction / degradation in visual and audio channel and *the display was augmented with vibrotactile cues*. Moreover, reaction time of the participant was faster when vibrotactile cues was involved as compared to when vibrotactile cues were not involved. Sensory modality successfully passed the significance level test. Our Null Hypothesis states that: there is no significant effect of sensory modality on task completion is false since the ($p < 0.000$). So, our Alternate Hypothesis: There is a significant effect of sensory modality on task completion. This significance level has been tested for ($\alpha < 0.05$) and we achieved a 95% confidence Interval (CI) for the effect of sensory modality on task completion.

Figure 25 shows the basic differences among the sensory modalities (visual, vibrotactile, audio) and their combinations as shown in Table IV. For task completion and reaction time, the sensory display involving vibrotactile cues produced better means than the sensory modalities that do not incorporate vibrotactile cues. Since, the experiment in which all the sensory modalities (vibrotactile + visual + audio) have been used, depicts the highest mean for task completion. So, it is evident that sensory modality / channel used for task completion has significant main effect for different sensory displays.

*B. Effect of Reaction Time on Task Completion*

Two hypotheses (Null and Alternate) have been formulated to asses whether there is an effect of Reaction Time on Task Completion or not. Null Hypothesis states that there is no significant effect of reaction time on task completion if ($p > 0.05$) for 95% Confidence Interval (CI). Alternate Hypothesis states that there is significant effect of reaction time on task completion if $p < 0.05$ for 95% Confidence Interval (CI). We performed three-way ANOVA and found out in the results that $p < 0.000$ in Table V which means that our null hypothesis is false and alternate hypothesis is true. So, our claim of alternate hypothesis is verified, which states that there is significant effect of reaction time on task completion. The significance level / confidence interval (CI) was $\alpha < 0.05$.

Reaction time had significant main effect on the level of success of task completion. It is evident from the result in the Figure 26 that participants showed a very quick reaction to multi-modal display when vibrotactile cues were involved while performing the tasks. In multi-modal display if one channel has been distracted or degraded the rest of the channels were still able to provide the useful information to the participant about task completion. Especially the vibrotactile channel augmented the other distracted/degraded sensory channels. It has been observed that when visual and audio channels were distracted / degraded vibrotactile cues helped the user in completion of the task and improved the reaction time of participant for a certain navigation cue. Moreover, vibrotactile only and combination (vibro + visual + audio) gave almost similar results and their mean scores lie almost in the same region. So, we can say that when sensory channels are degraded / distracted or not present, vibrotactile only cues can help the user in performing the navigation task with promising results and without distracting / degrading the other sensory channels. Moreover, it has been observed in the results ( Figure 27) that fast reaction time is associated with more number of task completed and slow reaction time is associated with less number of completed tasks or less number of check points crossed.

| Seven Display Conditions | Modes | Vibrotactile Device | Visual Display | Audio Commands |
|---|---|---|---|---|
| Vibrotactile cues only | Single modality | ON | OFF | OFF |
| Visual cues only | Single modality | OFF | ON | OFF |
| Audio cues only | Single modality | OFF | OFF | ON |
| Vibrotactile+Visual cues | Bimodal | ON | ON | OFF |
| Vibrotactile+Audio Cues | Bimodal | ON | OFF | ON |
| Visual+Audio | Bimodal | OFF | ON | ON |
| Vibrotactile+Audio+Visual Cues | Multi-modal | ON | ON | ON |

Table IV: Combinations of sensory modalities used during ground navigation experiment

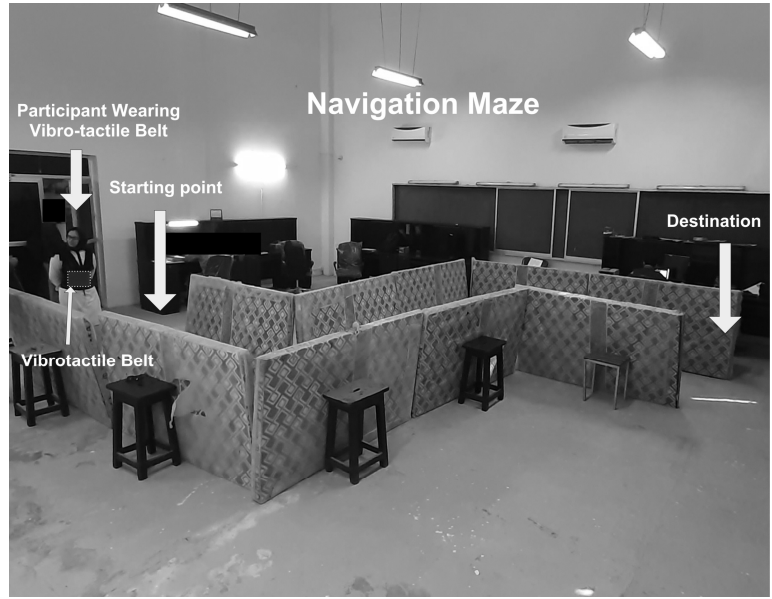

Figure 23: Participant wearing vibrotactile belt in navigation maze scenario while goal of the experiment is to guide the participant to walk through the maze from starting point till end point by taking vibrotactile cues from belt.

| Main Effect | F | d.f | p |
|---|---|---|---|
| Sensory Modality | 7.770 | 6 | 0.000 |
| Reaction Time | 172.729 | 3 | 0.000 |
| Gender | 0.651 | 1 | 0.421 |

Table V: Significant main effects from three-way ANOVA

### C. Effect of Gender on Task Completion

Two hypothesis (Null and Alternate) have been formulated for the main effect of gender to asses whether it has significant effect on task completion or not. Null hypothesis states that Gender (male, female) does not have any effect on task completion. Null hypothesis is considered to be true if $p > 0.05$ for Confidence Interval (CI) $\alpha < 0.05$. While alternate hypothesis states that gender has significant effect on task completion if $(p < 0.05)$ for Confidence Interval (CI) $\alpha < 0.05$. From the results of three-way ANOVA in Table V it has been found out that Gender (male, female) does not have significant effect on task completion so our Null Hypothesis is true because $(p > 0.05)$ if $(p > 0.05)$. In light of the results, it can be said that gender is not the decisive factor for task completion and device is equally suitable for both the genders.

### VII. CONCLUSIONS

Multiple user studies have been conducted on the proposed device to assess usability and effectiveness of a torso-mounted haptic display in enhancing situational awareness. Users were given a task of navigating through an unknown maze, while directional cues through different combinations of sensory modalities (vibrotactile, visual, audio) were provided to them. An assessment was made based on the successful navigation of maze checkpoints. With experimental data it has been observed that task completion has been increased when vibrotactile cues were involved no matter alone or in combination with other displays (visual and audio). Task completed was observed to be the highest for multi-modal display (visual + audio + vibrotactile). It was also observed that gender (male/female) does not present any significant effect on task completion. We are continuing this work with the aim of developing a viable vibrotactile display for enhancing situational awareness of vehicle drivers, aircraft pilots and for ground navigation.

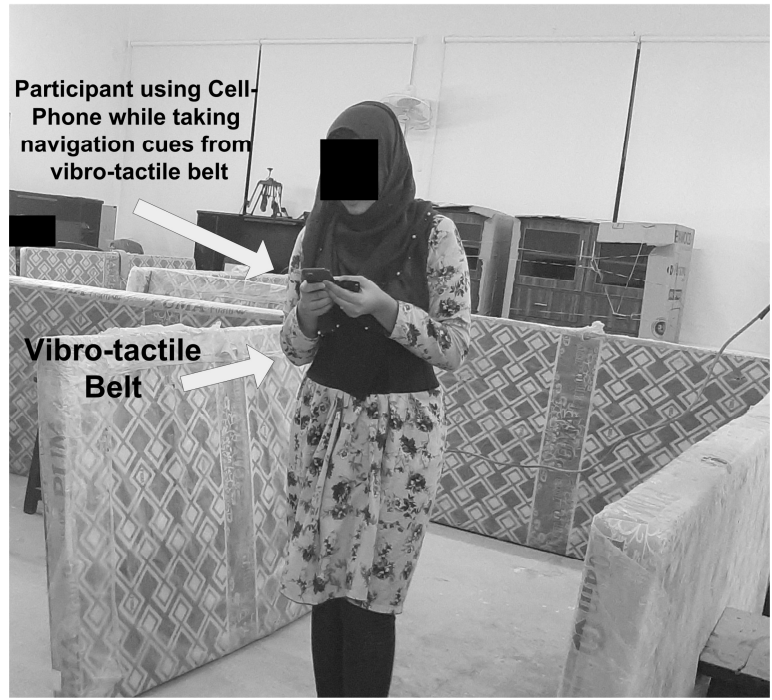

Figure 24: Participant attending to visual display of cell-phone screen, taking the audio commands / white noise from headphones and taking vibrotactile cues from belt.

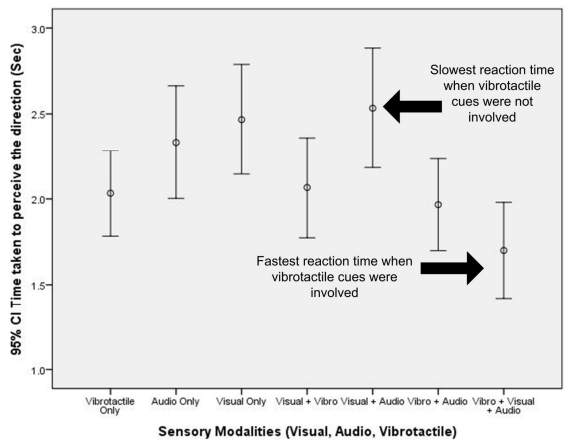

Figure 25: Reaction time (time taken to perceive the direction) of the participant varied with the change of Sensory Modalities

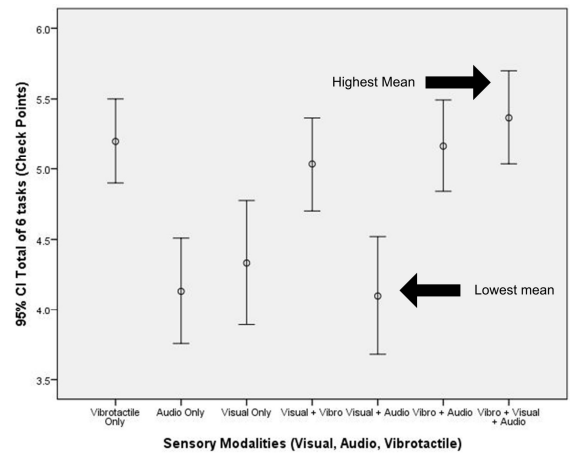

Figure 26: Task completion (check points crossed) has been improved when vibrotactile cues were used independently or in combination with other sensory modalities

## VIII. Acknowledgment

This work was partially supported by National Grassroots ICT Research Initiative.

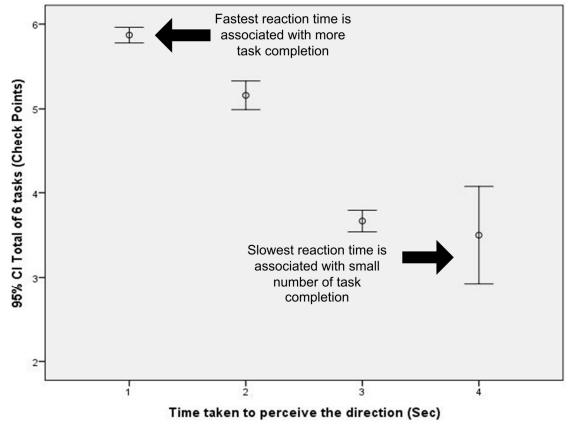

Figure 27: Fast reaction time is associated with more number of tasks completed and slow reaction time is associated with less number of completed tasks.

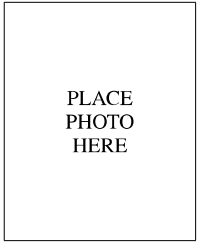

**Hammad Munawar** received his MS Control Systems Engineering in 2011 and PhD in 2017. He is currently an Assistant Professor at the Department of Avionics Engineering, National University of Sciences and Technology, Islamabad, Pakistan. His research interests include Control Systems, Robotics and Haptic Systems.

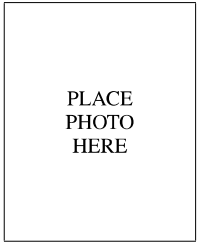

**Qasim Ali** received his MS Control Systems Engineering in 2010 and PhD in 2016. He is currently working as Head of Department (Research) at College of Aeronautical Engineering, National University of Sciences and Technology, Pakistan. His research interests include Applied Control Systems and Distributed Control.

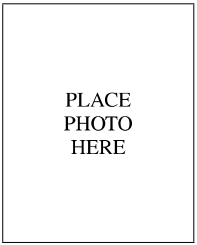

**Aakash Khaliq** is currently pursuing his Masters studies at the Department of Avionics Engineering, National University of Sciences and Technology, Islamabad, Pakistan. His research interests include Haptics and Control Systems.